\documentclass[preprint,nofootinbib,amsmath,amssymb,superscriptaddress]{revtex4-2}

\usepackage{graphicx}
\usepackage[colorlinks,linkcolor=magenta,anchorcolor=cyan,citecolor=blue]{hyperref}

\def\be{\begin{equation}}
\def\ee{\end{equation}}
\def\ba{\begin{eqnarray}}
\def\ea{\end{eqnarray}}

\def\lf{\left}
\def\rt{\right}

\begin{document}

\title{Broken blue-tilted inflationary gravitational waves: a joint analysis of NANOGrav 15-year and BICEP/Keck 2018 data}

\author{Jun-Qian Jiang}
\email{jiangjq2000@gmail.com}
\affiliation{School of Fundamental Physics and Mathematical
    Sciences, Hangzhou Institute for Advanced Study, UCAS, Hangzhou
    310024, China}
\affiliation{School of Physical Sciences, University of
Chinese Academy of Sciences, Beijing 100049, China}

\author{Yong Cai}
\email{caiyong@zzu.edu.cn}
\affiliation{School of Physics and Microelectronics, Zhengzhou University, Zhengzhou, Henan 450001, China}

\author{Gen Ye}
\email{ye@lorentz.leidenuniv.nl}
\affiliation{Leiden University, Instituut-Lorentz for Theoretical Physics, 2333CA, Leiden, Netherlands}

\author{Yun-Song Piao}
\email{yspiao@ucas.ac.cn}
\affiliation{School of Fundamental Physics and Mathematical
    Sciences, Hangzhou Institute for Advanced Study, UCAS, Hangzhou
    310024, China}
\affiliation{School of Physical Sciences, University of
Chinese Academy of Sciences, Beijing 100049, China}
\affiliation{International Center for Theoretical Physics
    Asia-Pacific, Beijing/Hangzhou, China}
\affiliation{Institute of Theoretical Physics, Chinese
    Academy of Sciences, P.O. Box 2735, Beijing 100190, China}

\begin{abstract}

Recently, the pulsar timing array (PTA) collaborations have
reported the evidence for a stochastic gravitational wave
background (SGWB) at nano-Hertz band. The spectrum of inflationary
gravitational wave (IGW) is unknown, which might exhibit different
power law at different frequency-bands, thus if the PTA signal is
primordial, it will be significant to explore the underlying
implications of current PTA and CMB data on IGW. In this Letter,
we perform a joint Markov Chain Monte Carlo analysis for a broken
power-law spectrum of IGW with the NANOGrav 15-year and BICEP/Keck
2018 data. It is found that though the bestfit spectral tilt of
IGW at PTA band is $n^\text{PTA}_\text{T} =2.42^{+0.32}_{-0.91}$,
at CMB band the bestfit is $n^\text{CMB}_\text{T}
=0.55^{+0.37}_{-0.10}$ while a detectable amplitude of $r$ with
$n^\text{CMB}_\text{T} \simeq 0$ is still compatible. The
implication of our results for inflation is also discussed.

\end{abstract}

\maketitle
\newpage

\section{Introduction}

It is well-known that the detection of inflationary or primordial
gravitational waves (IGW) will not only solidify our confidence on
the inflation scenario
\cite{Guth:1980zm,Linde:1981mu,Albrecht:1982wi,Starobinsky:1980te,Linde:1983gd},
but also bring us significant insight into the physics of very
early universe.

The ultra-low-frequency IGW with $f\sim 10^{-18}-10^{-16}$Hz will
source the B-mode polarization in the cosmic microwave background
(CMB) \cite{Kamionkowski:1996ks,Kamionkowski:1996zd}, which has
been still searched for by BICEP/Keck \cite{BICEP:2021xfz}.
Recently, the PTA experiments
\cite{NANOGrav:2023gor,Antoniadis:2023rey,Reardon:2023gzh,Xu:2023wog},
have found
a stochastic GW background (SGWB) at $f\sim 10^{-10}-10^{-8}$Hz
\cite{NANOGrav:2023hvm,EPTA:2023xxk}. Though such a SGWB is
compatible with that brought by inspiralling supermassive black
holes binaries
\cite{NANOGrav:2023hvm,EPTA:2023xxk,Broadhurst:2023tus}, see also
\cite{Huang:2023chx,Depta:2023qst} for the supermassive primordial
black holes, it might be just IGW but with $n_\text{T}>0$
\cite{NANOGrav:2023hvm,EPTA:2023xxk,Vagnozzi:2020gtf,Vagnozzi:2023lwo,Ben-Dayan:2023lwd}\footnote{In
addition, see also recent other possibilities
e.g.\cite{Madge:2023cak,Franciolini:2023pbf,Ghosh:2023aum,Niu:2023bsr,Konoplya:2023fmh,Li:2023bxy,DiBari:2023upq,Zhu:2023faa,Du:2023qvj,Ye:2023xyr,Balaji:2023ehk,Zhang:2023nrs,Oikonomou:2023qfz,Madge:2023cak,Franciolini:2023pbf,Ashoorioon:2022raz,Cheung:2023ihl,Li:2023tdx,HosseiniMansoori:2023mqh}.}.

However, the spectrum of IGW is actually unknown. The amplitude of
IGW at CMB band is usually quantified as the tensor-to-scalar
ratio $r={A_\text{T}\over A_s}$, we have
\begin{equation}
P_\text{T}(k) = r A_s
\left(\frac{k}{k_\text{pivot}}\right)^{n_\text{T}},
\label{PT1}\end{equation} where $k_\text{pivot}$ is the pivot
scale. It is possible for inflation to yield a blue-tilted SGWB
with $n_\text{T}>0$
\cite{Piao:2004tq,Piao:2003ty,Piao:2004jg,Baldi:2005gk,Liu:2010dh,Kobayashi:2010cm,Kobayashi:2011nu,Fujita:2018ehq,Endlich:2012pz,Cannone:2014uqa,Giare:2020plo,Cai:2016ldn,Cai:2015yza,Baumgart:2021ptt,Ben-Dayan:2023lwd},
however, the primordial scalar perturbation is nearly
scale-invariant at CMB band \cite{Planck:2018vyg}, which seems to
be in favor of a slow-roll model of inflation with
$n_\text{T}\simeq 0$ \cite{Planck:2018jri}. Inspired by
\cite{Cai:2020qpu,Tahara:2020fmn}, see also
\cite{Liu:2011ns,Liu:2012ww,Nishi:2015pta,Nishi:2016ljg,Giare:2022wxq},
it might be more reliable to consider a broken power-law IGWB at
$f\sim 10^{-18}-10^{-8}$Hz,
\begin{equation}
P_\text{T}(k) = r A_s
\left(\frac{k}{k_\text{pivot}}\right)^{n^\text{CMB}_\text{T}}
\left(1 +
\lf(\frac{k}{k_\text{break}}\rt)\right)^{-n^\text{CMB}_\text{T}+n^\text{PTA}_\text{T}}
, \label{PT2}\end{equation}
where
$f_\text{break}\ll 10^{-8}$Hz is the scale at which power-law is
broken,
and the conversion between $k$ and $f$ is
\begin{equation}
    f = \frac{k c }{2 \pi a_0} = 1.5 \times 10^{-15} \left( \frac{k}{\text{Mpc}^{-1}} \right) \text{Hz} .
\end{equation}
In
Refs.\cite{Kuroyanagi:2014nba,Kuroyanagi:2020sfw,Benetti:2021uea}
such a broken SGWB has been studied but with
$f_\text{break}>10^{-7}$Hz. According to \autoref{PT2}, we have
$P_\text{T}= r A_s
\left(\frac{k}{k_\text{pivot}}\right)^{n^\text{CMB}_\text{T}}$
when $k \ll k_\text{break}$, while \be P_\text{T} = r A_s
\left(\frac{k_\text{break}^{n^\text{CMB}_\text{T} -
n^\text{PTA}_\text{T}}}{k_\text{pivot}^{n^\text{CMB}_\text{T}}}\right)
k^{n^\text{PTA}_\text{T}},\label{PT3}\ee when $k \gg
k_\text{break}$. Actually, a period of inflation might be complex
so that at different bands of $f\sim 10^{-18}-10^{-8}$Hz we will
have different power-law IGW, while \autoref{PT2} is just the
simplest of such possibilities \footnote{Actually, beyond PTA band
such an IGW spectrum will be conflicted with the BBN bound on
relativistic components, however, we only consider (\ref{PT2}) at
$f\sim 10^{-18}-10^{-8}$Hz, since at higher-frequency band
(\ref{PT2}) might have been modified
\cite{Kuroyanagi:2014nba,Kuroyanagi:2020sfw,Benetti:2021uea}, see
also section-III.}.

It has been found that for power-law IGW (\ref{PT1}), recent
NANOGrav data favors $n_\text{T}\simeq 2$. As a result, IGW at CMB
band is negligible. Thus recent CMB data has not been included in
the analysis of NANOGrav \cite{NANOGrav:2023hvm} (a hard prior
$r\lesssim 0.03$ is actually sufficient \cite{Vagnozzi:2023lwo}).
However, this is not valid for broken power-law IGW (\ref{PT2}),
which allows a non-negligible $r$ at CMB band. Thus it is
significant to perform a joint analysis of both latest PTA and CMB
data with full exact likelihood to explore the underlying impact
of current data on IGW.

However, a joint Markov Chain Monte Carlo (MCMC) analysis of
NANOGrav 15-year and BICEP/Keck 2018 data has been still open. In
this Letter, we present the first such analysis, and find that the
bestfit spectral tilt of IGW at PTA band is $n^\text{PTA}_\text{T}
\simeq 2$, however, different from the results of NANOGrav
\cite{NANOGrav:2023hvm}, $n^\text{CMB}_\text{T} \simeq 2$ is not
favored at CMB band, instead the bestfit is $n^\text{CMB}_\text{T}
\simeq 0.5$ while a detectable amplitude of $r$ with
$n^\text{CMB}_\text{T} \simeq 0$ is still compatible. The
implication of our results for inflation is also discussed.

\section{Dataset and Method}

\textbf{NANOGrav}: We use the NANOGrav 15-year dataset
\cite{NANOGrav:2023gor} at PTA band, assuming that the signals
observed in NANOGrav \cite{NANOGrav:2023gor}, EPTA
\cite{Antoniadis:2023rey}, PPTA \cite{Reardon:2023gzh} and CPTA
\cite{Xu:2023wog} are mutually consistent.
The likelihoods are calculated with \texttt{ceffyl}
\cite{Lamb:2023jls}.

\textbf{BICEP/Keck (BK18)}: We use the BICEP/Keck 2018 official
likelihood\footnote{\url{http://bicepkeck.org/bk18_2021_release.html}}
\cite{BICEP:2021xfz} at CMB band, taking dust, synchrotron and
noise into account.

As in the analysis of NANOGrav \cite{NANOGrav:2023hvm}, we fix the
parameters of standard $\Lambda$CDM model to the bestfit values of
the Planck 2018 baseline results:\footnote{The corresponding
bestfit values might shift in light of the early resolution of the
Hubble tension \cite{Ye:2021nej}, in particular the spectral index
$n_s$ of primordial scalar perturbation will shift towards $n_s=1$
\cite{Ye:2020btb,Jiang:2022uyg,Jiang:2022qlj}, however, such
shifts will not essentially alter our results. } $\Omega_b
h^2=0.02238$, $\Omega_c h^2=0.12011$, $100\theta_{MC}=1.040909$,
$\tau=0.0543$, $\ln(10^{10}A_s)=3.0448$, $n_s=0.96605$. In
addition to the nuisance parameters in the BICEP/Keck likelihood,
our MCMC parameters set is $\{r_{0.05}, n^\text{CMB}_\text{T},
n^\text{PTA}_\text{T}, \log_{10}{k_\text{break}}\}$, where the
subscript $0.05$ for $r$ indicates that it is calculated at the
pivot scale $k=0.05$Mpc$^{-1}$. The uniform priors are set, and the
unit of $k_\text{break}$ is Mpc$^{-1}$.

And we use \texttt{CLASS} \cite{Blas:2011rf} to calculate the
evolutions of GWs and other components, such as photons and
baryons, and \texttt{cobaya} \cite{Torrado:2020dgo} with the MCMC
Metropolis sampler and oversampling the nuisance parameters and
$\{n_\text{T}^\text{PTA}, k_{break}\}$ to speed up our calculation.



Here, we also need to calculate the energy spectrum
$\Omega_\text{GW}(f)$ of IGW. As in the analysis of NANOGrav
\cite{NANOGrav:2023hvm}, around the PTA scales we have
\begin{equation} \label{eq:energy_spectrum}
\Omega_\text{GW}(f) = \frac{\Omega_\text{r}}{24} \left(
\frac{g_*(f)}{g_*^0} \right) \left( \frac{g_{*,s}^0}{g_{*,s}(f)}
\right)^{4/3} P_\text{T}(f),
\end{equation}
where $\Omega_\text{r}$ is the ratio of radiation to critical
energy density at present, and $g_*^0$ and $g_{*,s}^0$ are the
effective number of relativistic degrees of freedom for energy and
entropy, respectively, while $g_*(f)$ and $g_{*,s}(f)$ correspond
to the qualities when the modes with comoving wavenumber $k=2\pi
a_0 f$ reentered the Hubble horizon, see
e.g. Ref.\cite{Saikawa:2020swg}.


\section{Results}
\label{sec:Results}

In this section, we will focus on the broken power-law model
(\ref{PT2}) of IGW with $n^\text{CMB}_\text{T}$ free and
$n^\text{CMB}_\text{T} = - r/8$, respectively.

\subsection{$n^\text{CMB}_\text{T}$ is free}

\begin{figure}[h!]
\centering
\includegraphics[width=0.9\textwidth]{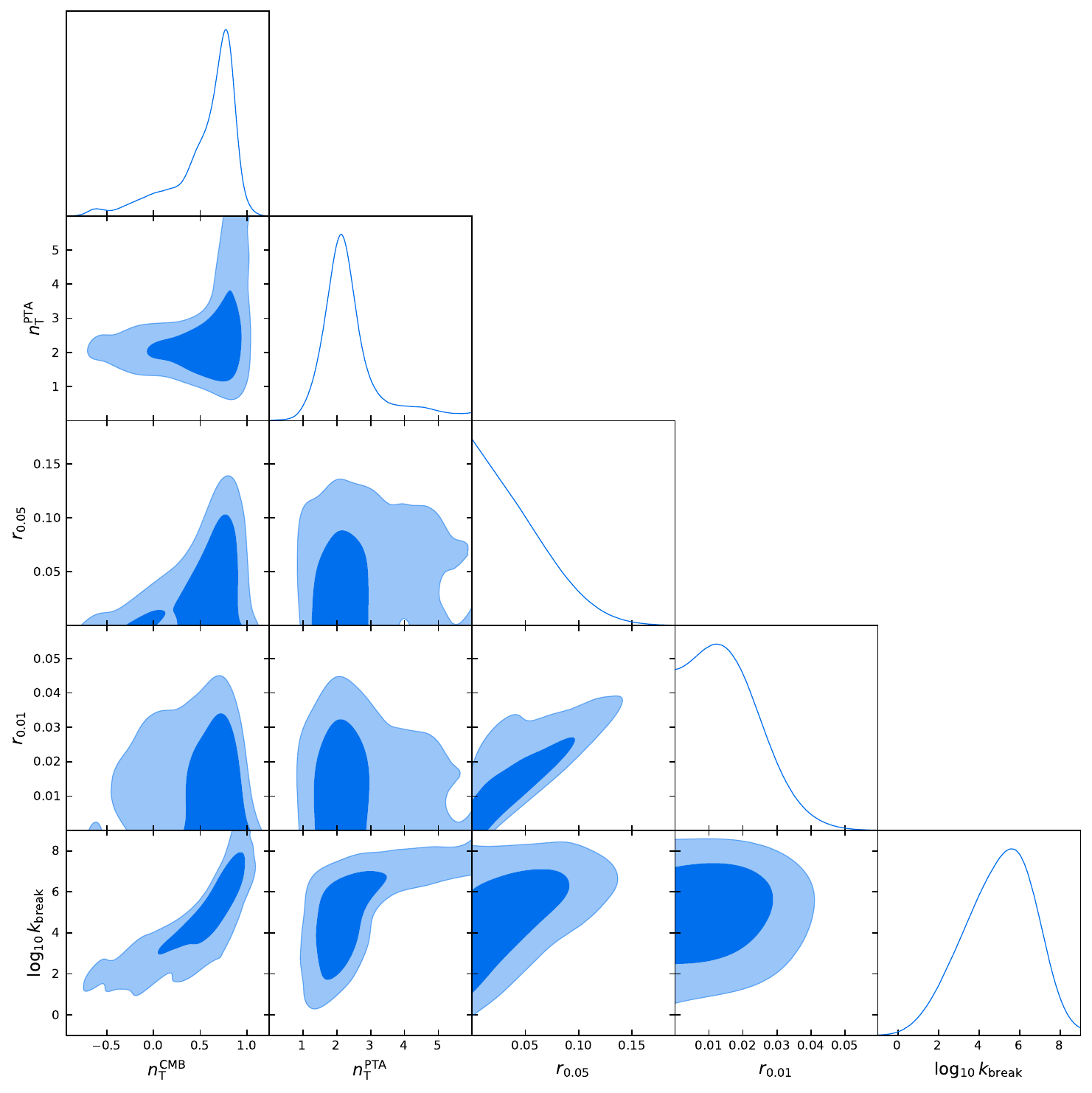}
\caption{\label{fig:broken_free_nt}Marginalized posterior
distributions of the parameters in the broken power-law model
(\ref{PT2}) of IGW with $n_\text{T,CMB}$ free.}
\end{figure}

\begin{table}[h!]
\centering
\begin{tabular} {c|cl}
 Parameter &  Best fit &  68\% limits\\
\hline
$r_{0.05}       $ & $0.0296                    $ & $< 0.0549 \, (<0.107)              $\\
$r_{0.01}       $ & $0.0141                    $ & $< 0.0199 \, (<0.0338)              $\\
$n^\mathrm{CMB}_\mathrm{T}$ & $0.428                     $ & $0.55^{+0.37}_{-0.10}      $\\
$n^\mathrm{PTA}_\mathrm{T}$ & $2.08                      $ & $2.42^{+0.32}_{-0.91}      $\\
$\log_{10}{k_\mathrm{break}}$ & $4.19                      $ & $5.0^{+2.0}_{-1.5}         $\\
\end{tabular}
\caption{The bestfit values and the 68\% limit of posterior
distributions of the parameters in the broken power-law model
(\ref{PT2}) of IGW with $n_\text{T,CMB}$ free.
We also show the 95\% limit for $r$ in parentheses.}
\label{tab:broken_free_nt}
\end{table}

The results are presented in \autoref{fig:broken_free_nt} and
\autoref{tab:broken_free_nt}. As expected, the bestfit value
$\log_{10} k_\mathrm{break}=4.19$, which is just between the CMB
and PTA bands. The posterior of the spectral index
$n^\text{PTA}_\text{T}$ at PTA band is $n^\text{PTA}_\text{T} \simeq 2$, in
agreement with the results of NANOGrav
\cite{NANOGrav:2023hvm,Vagnozzi:2023lwo}\footnote{Here, our
results corresponds to the $T_\text{rh} \gtrsim 1$ GeV part of
Ref.\cite{NANOGrav:2023hvm} }, since it is constrained mainly by
the PTA itself. According to \autoref{PT3}, we have $P_\text{T} \sim r
k^{n^\text{PTA}_\text{T}}
k_\text{break}^{n^\text{CMB}_\text{T}-n^\text{PTA}_\text{T}}$ for fixed
$k_\text{pivot}$, which suggests that $n^\text{CMB}_\text{T}$ is
correlated positively with $\log_{10}{k_\text{break}}$, as showed
in \autoref{fig:broken_free_nt}.

The 95\% upper limit of $r_{0.05}$ is $r_{0.05}< 0.107$, higher
than that of Planck+BK18 (e.g. $r\lesssim 0.03$ in
Refs.\cite{BICEP:2021xfz,Tristram:2021tvh}), while the 95\% upper
limit of $r_{0.01}$ is $r_{0.01}< 0.0338$, However, this is a
natural result, since for a blue-tilted IGW with the bestfit
$n^\text{CMB}_\text{T} = 0.428$, the amplitude of IGW must be lower at
larger scale (smaller $k_\text{pivot}$). Accordingly, the
constraint will be notably different at different
$k_{pivot}\lesssim 0.05$ Mpc$^{-1}$.

However, it is significant to find that unlike that of NANOGrav
\cite{NANOGrav:2023hvm} at CMB band $n^\text{CMB}_\text{T} \simeq
2$ is not favored, instead $n^\text{CMB}_\text{T} =
0.55^{+0.37}_{-0.10}$ with the bestfit $n^\text{CMB}_\text{T} =
0.428$, while $n^\text{CMB}_\text{T} \simeq 0$ is still in the
$95\%$ CL. range, which suggests that a slow-roll period of
inflation at CMB band is not excluded. Thus it is interesting to
see what if we fix $n^\text{CMB}_\text{T}=-r/8$.

\subsection{$n^\text{CMB}_\text{T}=-r/8$}

\begin{figure}[h!]
\centering
\includegraphics[width=0.65\textwidth]{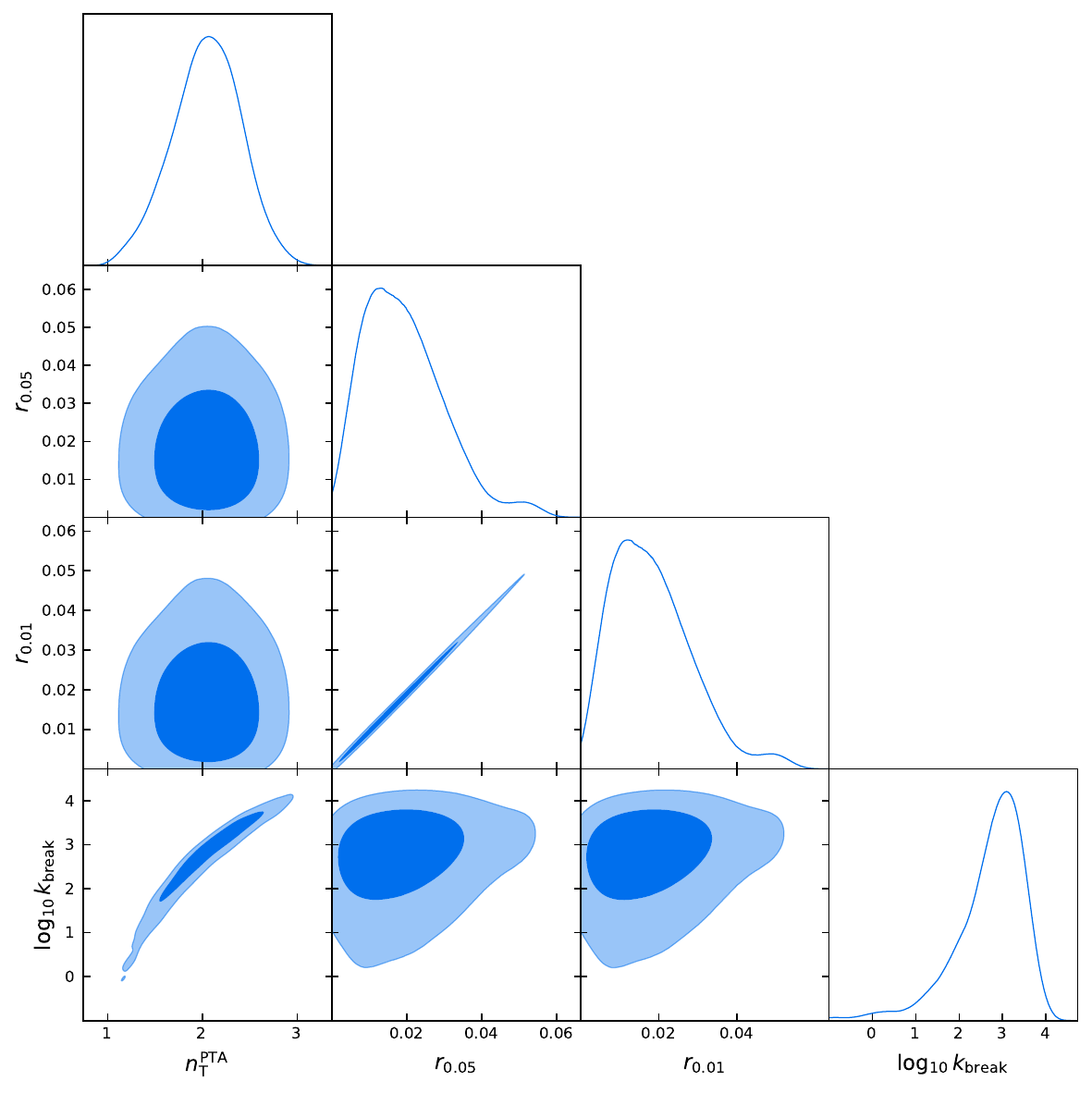}
\caption{\label{fig:broken_scc_nt}Marginalized posterior
distributions of the parameters in the broken power-law model
(\ref{PT2}) with slow-roll $n_\text{T,CMB}$.}
\end{figure}

\begin{table}[h!]
\centering
\begin{tabular} {c|cl}
 Parameter &  Best fit &  68\% limits\\
\hline
$r_{0.05}       $ & $0.0091                    $ & $0.0194^{+0.0069}_{-0.014} \, (^{+0.020}_{-0.019}) $\\
$r_{0.01}       $ & $0.0086                    $ & $0.0184^{+0.0065}_{-0.013} \, (^{+0.019}_{-0.018}) $\\
$n^\mathrm{PTA}_\mathrm{T}$ & $2.198                     $ & $2.04^{+0.39}_{-0.34}      $\\
$\log_{10}{k_\mathrm{break}}$ & $3.01                      $ & $2.72^{+0.92}_{-0.40}      $\\
\end{tabular}
\caption{The bestfit values and the 68\% limit of posterior
distributions of the parameters of the broken power-law model
(\ref{PT2}) with slow-roll $n_\text{T,CMB}$.
We also show the 95\% limit for $r$ in parentheses.}
\label{tab:broken_scc_nt}
\end{table}

Next, we fix $n_\text{T,CMB} = - r / 8$, i.e. standard slow-roll
inflation is not broken at CMB scale. The results are presented in
\autoref{fig:broken_scc_nt} and \autoref{tab:broken_scc_nt}. As
expected, the broken scale is
$k_\text{break}=2.72^{+0.92}_{-0.40}$, which is well between the
CMB and PTA bands, and the posterior of the spectral index
$n^\text{PTA}_\text{T}$ at PTA band is still $n^\text{PTA}_\text{T} \simeq
2$.

The 95\% CL. range of $r$ is $r_{0.05}=0.019^{+0.020}_{-0.019}
$, which is similar to that of $r_{0.01}$, since unlike in
\autoref{sec:Results}.A, we have $n^\text{CMB}_\text{T}=-r/8\simeq 0$
nearly scale-invariant. This upper limit, $r\lesssim 0.039$, is
comparable with (but slightly higher than) that of Planck+BK18
\cite{BICEP:2021xfz,Tristram:2021tvh}. The bestfit $r$ is
$r_{0.05}=0.0091$, however, it is unexpected that $r=0$ is at $2\sigma$
level. According to \autoref{PT2}, the bound on $r$ seems to be
indirectly affect the NANOGrav results ($n^\text{PTA}_\text{T}$,
$k_\text{break}$).

\section{Discussion}
\label{sec:Discussion}

\subsection{Conclusion}

\begin{figure}[h!]
\centering
\includegraphics[width=0.7\textwidth]{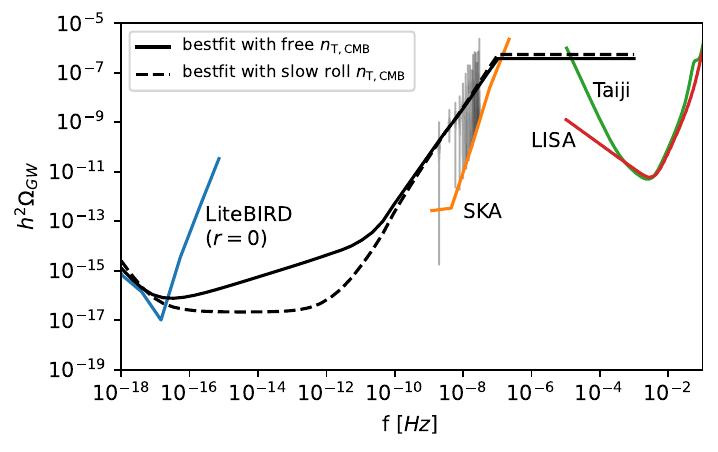}
\caption{\label{fig:future} The bestfit energy spectrum
$\Omega_{GW}(f)$ of IGWB with $n_\text{T}^\text{CMB}$ free and
$n_\text{T}^\text{CMB}=-r/8$, respectively. We also show the
sensitivity curves for LiteBIRD, SKA (Square Kilometre Array), space-based laser
interferometers: LISA and Taiji. Here, we use the transfer
function showed in Ref.\cite{Kite:2021yoe}, which can return to
\autoref{eq:energy_spectrum} at the high frequency limit, see also
Refs.\cite{Turner:1993vb,Boyle:2005se,Zhao:2006mm,Kuroyanagi:2014nba,Liu:2015psa}.
As commented in \autoref{sec:Discussion}, when $f>10^{-7}$ the
highly-blue ($n_\text{T}^\text{PTA}\simeq 2$) tilt needs to be cut off, though
we plot $\Omega_{GW}(f>10^{-7})$ with a straight line, which but
is actually model-dependent,
e.g.\cite{Kuroyanagi:2020sfw,Benetti:2021uea}. }
\end{figure}

In conclusion, we find that though the bestfit spectral tilt of
IGW at PTA band is $n^\text{PTA}_\text{T} \simeq 2$, unlike that
of the NANOGrav \cite{NANOGrav:2023hvm} $n^\text{CMB}_\text{T}
\simeq 2$ is not favored at CMB band, instead the bestfit is
$n^\text{CMB}_\text{T} =0.55^{+0.37}_{-0.10}$, while a detectable
amplitude of $r$ with $n^\text{CMB}_\text{T}\simeq 0$ is still
compatible. In particular if we fix $n^\text{CMB}_\text{T}=-r/8$
in light of standard slow-roll inflation, the upper bound on $r$
is $r\lesssim 0.039$, slightly higher than that of Planck+BK18
\cite{BICEP:2021xfz}

In \autoref{fig:future}, we show our bestfit energy spectrum
$\Omega_{GW}(f)$ of IGWB with $n_\text{T}^\text{CMB}$ free and
$n_\text{T}^\text{CMB}=-r/8$, respectively. The sensitivity curves
for LiteBIRD and SKA are based on the results of
Ref.\cite{Campeti:2020xwn}. Here, our bestfit $\Omega_{GW}(f)$
covers the band at $f\sim 10^{-18}-10^{-8}$Hz, which naturally
offers a guide for building relevant models of inflation and
exploring new physics at corresponding band, since such a bestfit
$\Omega_{GW}(f)$ can simultaneously explain current observations
at both CMB and PTA bands.

{ 
The GW energy density contributes to the effective number of relativistic degrees of freedom $N_\text{eff}$ at the time of BBN, which can be constrained by observations (see e.g. \cite{Meerburg:2015zua,Ben-Dayan:2019gll}):
\begin{equation}
    N_\text{eff}^\text{GW} = \left( 3.046 + \frac{8}{7} \left( \frac{11}{4} \right) ^{4/3} \right) \frac{1}{12} \int \mathrm{d} \ln k P_T
\end{equation}
The IR cutoff for the integral is the comoving horizon at the time of BBN $f_\text{IR} \simeq 10^{-10}$ Hz.
Our expression of the power spectrum is assumed to be valid until $f=2.97\times 10^{-8}$ Hz,
which is the frequency of the 14th bin of the NANOGrav 15-yr result.
We find a contribution of $\Delta N_\text{eff} = 0.012$ for a free $n_\text{T}^\text{CMB}$ and $\Delta N_\text{eff} = 0.014$ for $n_\text{T}^\text{CMB} = -r/8$,
which is consistent will current constraint $\Delta N_\text{eff} \lesssim 0.4$
(95\% CL., see e.g. \cite{Planck:2018vyg}).
}

However, beyond the PTA band the highly-blue
($n_\text{T}^\text{PTA}\simeq 2$) tilt must be cut off at a certain
$k_{cutoff}=2\pi f_{cutoff}$ to avoid the confliction with the BBN
bound.
{ 
For example, if the power spectrum maintains flat till some $f_\text{UV} \sim 10^8$
\footnote{The choice of $f_\text{UV}$ has many assumptions. Here we consider the value in Refs.\cite{Meerburg:2015zua,Cabass:2015jwe}.}
when $f > 10^{-7}$, it will lead to $\Delta N_\text{eff} = 1.01$ for a free $n_\text{T}^\text{CMB}$ and $\Delta N_\text{eff} = 1.27$ for $n_\text{T}^\text{CMB} = -r/8$.
We find a cutoff frequency $f_\text{cutoff} \simeq 10^{-3}$ can produce $\Delta N_\text{eff} \simeq 0.4$, which has not been excluded by the current constraint.
Besides, if the power spectrum is kept flat till $f=$20-76.6 Hz,
the energy density of the GWs ($3.72\times 10^{-7}$ for a free $n_\text{T}^\text{CMB}$ and $5.45\times 10^{-7}$ for $n_\text{T}^\text{CMB} = -r/8$) will violate the upper limit of the LIGO/Virgo constraint $\Omega_\text{GW} \lesssim 5.8\times 10^{-9}$ in the band 20-76.6 Hz \cite{KAGRA:2021kbb}.
It also required the cutoff to not conflict with these results. 
}
In \autoref{fig:future}, what is $P_\text{T}$ at
$k>k_{cutoff}$ is not relevant with our MCMC results, which
actually is model-dependent,
e.g.\cite{Kuroyanagi:2020sfw,Benetti:2021uea}, see also discussion
in \autoref{sec:Discussion}.B. Thus the spectrum of IGWB at that
band is open, however, the space-based laser interferometers, LISA
and Taiji, might detect the corresponding IGWB signal.

In next decade, with the accumulations of PTA and CMB data, SGWB
at PTA band would be confirmed eventually, and also upcoming CMB
observations, such as CMB-S4 \cite{CMB-S4:2020lpa}, LiteBIRD
\cite{Hazumi:2019lys} would bring us more information on IGWB at
CMB band. In view of this, our work is suggesting that a joint
MCMC analysis of PTA and CMB data might be indispensable, which
would open a unforeseen door into our very early Universe.

\subsection{Implications for inflation}

It is interesting to see what our results will imply for
inflation. As commented in Ref.\cite{Vagnozzi:2023lwo}, a
blue-tilted IGW suggests that inflation must violate null energy
condition (NEC), i.e. $T_{\mu\nu}n^{\mu}n^{\nu}<0$ (equivalently
${\dot H}>0$ or $\epsilon=-{{\dot H}\over H^2}<0$),
e.g.\cite{Piao:2004tq,Piao:2003ty}, see also
\cite{Kobayashi:2010cm,Kobayashi:2011nu}, if the initial
Bunch-Davis state of perturbation modes is not modified. Though it
is not difficult to achieve $n^\text{PTA}_\text{T} \simeq 2$,
which requires $\epsilon\ll -1$,
e.g.\cite{Piao:2004jg,Liu:2011ns}, such a highly blue tilt of IGWB
implies that inflation must end at a low scale
\cite{NANOGrav:2023hvm,Vagnozzi:2023lwo}, or else it will be
conflicted with the BBN bound on relativistic components.

However, it might be possible that the violation of NEC happened
before a (standard) slow-roll inflation, so that the blue-tilted
$P_\text{T}$ is cut off at certain $k_{cutoff}$ beyond the PTA
band (when $k>k_{cutoff}$ the spectrum is flat). In corresponding
models, the energy spectrum of IGW will be shown as the black
solid $\Omega_{GW}(f)$ curve in \autoref{fig:future}, and at CMB
band for such a period of NEC violation the scale-invariant
primordial scalar perturbation required by Planck observations can
be harvested in light of
\cite{Piao:2004tq,Kobayashi:2010cm,Piao:2010bi,Liu:2011ns,Liu:2012ww,Nishi:2015pta,Nishi:2016ljg}.

It is also possible that the violation of NEC might be just
short-lived, which happened between two (slow-roll) periods of
inflation with $H\simeq H_{inf1}$ and a higher scale $H\simeq
H_{inf2}\gg H_{inf1}$, respectively. In corresponding model
\cite{Cai:2020qpu,Cai:2022nqv,Cai:2023uhc}, we approximately have
\ba {P_\text{T}(k)} \sim {{{H}^2_{inf1}} + {\cal A}\lf({k\over
k_{cutoff}}\rt)^{n_\text{T}^\text{PTA}} H^2_{inf2}\over 1 +  {\cal
A}\lf({k\over k_{cutoff}}\rt)^{n_\text{T}^\text{PTA}} }\,,
\label{eq:para01} \ea where ${\cal A}={\Gamma^2(\nu)\over \pi}
\lf({2\nu-1 \over 4}\rt)^{1-2\nu}\sim {\cal O}(1)$ with
$\nu=1/2+{1\over 1-\epsilon}$, \be
n_\text{T}^\text{PTA}=-{2\epsilon \over 1-\epsilon}\simeq 2,\quad
for\quad \epsilon\ll 1.\ee According to \autoref{eq:para01}, we
have $P_\text{T}\simeq H_{inf1}^2$ for $k/k_{cutoff}\ll
({H_{inf1}\over H_{inf2}})^{2/n_\text{T}^\text{PTA}}\ll 1$
corresponding the CMB band, while $P_\text{T}\sim
k^{n_\text{T}^\text{PTA}}\simeq k^2$ for $({H_{inf1}\over
H_{inf2}})^{2/n_\text{T}^\text{PTA}}\ll k/k_{cutoff}\ll 1$
corresponding the PTA band. Thus when $k/k_{cutoff}\ll 1$, we have
\ba {P_\text{T}(k)} \sim {{H}^2_{inf1}} \lf[1+ {\cal A}\lf({k\over
k_{cutoff}}\rt)^{n_\text{T}^\text{PTA}} {H^2_{inf2}\over
{H}^2_{inf1}} \rt]\,. \label{eq:para02} \ea It is just
\autoref{PT2} for $k_{\rm break}\simeq k_{cutoff} ({H_{inf1}\over
H_{inf2}})^{2/n_\text{T}^\text{PTA}}$ and
$n_\text{T}^\text{CMB}\simeq 0$. Thus the energy spectrum will be
showed as the black dashed $\Omega_{GW}(f)$ curve in
\autoref{fig:future}. Here, $k_{cutoff}$ is a cutoff scale, see
also \autoref{sec:Discussion}.A, beyond which $P_\text{T}\sim
H_{inf2}^2$ is flat, or might be modified as in e.g.
Refs.\cite{Kuroyanagi:2014nba,Kuroyanagi:2020sfw,Benetti:2021uea}.
Thus the IGWB at $k>k_\text{cutoff}$ ($f>10^{-7}$Hz in
\autoref{fig:future}) is model-dependent, which is unknown and
needs to be explored, which suggests that the requirement of BBN
that inflation must end at a low scale
\cite{NANOGrav:2023hvm,Vagnozzi:2023lwo} is not workable.

It is also interesting to investigate other mechanisms, in which
blue-tilted IGW at PTA band is sourced by other components during
slow-roll inflation, such as the gauge fields
\cite{Anber:2012du,Cook:2011hg,Mukohyama:2014gba,Namba:2015gja,Dimastrogiovanni:2016fuu,Obata:2016oym,Caldwell:2017chz},
the non-Bunch-Davis initial states
\cite{Ashoorioon:2014nta,Choudhury:2023kam}, modified gravity,
e.g.\cite{Oikonomou:2022ijs,Oikonomou:2022xoq,Odintsov:2021kup,Odintsov:2022cbm},
and the collisions of bubbles nucleated
\cite{Li:2020cjj,Wang:2018caj}. In light of our MCMC results on
$r$, $n_\text{T}^\text{CMB}$ and $n_\text{T}^\text{PTA}$, and also
bestfit $\Omega_{GW}(f)$ in \autoref{fig:future}, it is
interesting to resurvey the relevant models.

\section*{Acknowledgments}
YSP is supported by NSFC, No.12075246 and the Fundamental Research
Funds for the Central Universities.
Y. C. is supported in part by the National Natural Science Foundation
of China (Grant No. 11905224), the China Postdoctoral Science
Foundation (Grant No. 2021M692942) and Zhengzhou University (Grant No. 32340282).
GY is supported by NWO and the Dutch
Ministry of Education, Culture and Science (OCW) (grant VI.Vidi.192.069).

\bibliography{refs}

\end{document}